%% file: main.tex
\documentclass[letterpaper, 10 pt, journal, twoside]{IEEEtran}

\IEEEoverridecommandlockouts                              

\usepackage{xr}

\newcommand{\omitted}[1]{}

\input{titleAndAuthors}

\usepackage{cite}

\usepackage{comment}
\usepackage{siunitx}
\usepackage{relsize}
\usepackage{ifthen}
\usepackage[colorinlistoftodos]{todonotes}

\usepackage[caption=false]{subfig}

\usepackage[vlined,ruled,linesnumbered]{algorithm2e}
\usepackage{graphics} 
\usepackage{rotating}
\usepackage{color}
\usepackage{enumerate}
\usepackage[T1]{fontenc}
\usepackage{psfrag}
\usepackage{epsfig} 
\usepackage{booktabs}
\usepackage{graphicx,url}
\usepackage{multirow}
\usepackage{array}
\usepackage{latexsym}
\usepackage{amsfonts}
\usepackage{amsmath}
\usepackage{amssymb}
\usepackage{xstring}
\usepackage{algorithmic}
\usepackage{multirow}
\usepackage{xcolor}
\usepackage{prettyref}
\usepackage{flexisym}
\usepackage{bigdelim}
\usepackage{breqn} 
\usepackage{listings}
\let\labelindent\relax
\usepackage{xspace}
\usepackage{bm}
\graphicspath{{./figures/}}
\usepackage{tikz}
\usetikzlibrary{matrix,calc}
\usepackage{lipsum}
\usepackage{mdwlist}

\let\itemize\undefined
\makecompactlist{itemize}{stditemize}
\usepackage{enumitem}
\usepackage{caption}
\usepackage{amsthm}
\usepackage{mathtools}

\makeatletter
\let\NAT@parse\undefined
\makeatother
\usepackage{hyperref}
\usepackage{cleveref}

\hypersetup{%
  colorlinks=true,
  linkcolor=blue,
  filecolor=magenta,      
  urlcolor=red,
  citecolor=red,
  linkbordercolor={0 0 1}
}

\input{preamble_symbols.tex}

\input{shortcuts.tex}

\PassOptionsToPackage{end}{algorithmic}

\renewcommand{\baselinestretch}{0.96}

\begin{document}

\maketitle

\input{Abstract.tex}

\input{1-Introduction.tex}

\input{2-Problem.tex} 

\input{3-Algorithms.tex} 

\input{4-Guarantees.tex} 

\input{5-Experiments.tex}

\input{6-Conclusion.tex}

\vspace{-2mm}
\appendices
\input{new-App.tex} 

\vspace{-4mm}
\bibliographystyle{IEEEtran}
\bibliography{references,security_references}

\end{document}

%% file: titleAndAuthors.tex
\title{
 \fontsize{19}{19} \selectfont 
Online Submodular Coordination with Bounded Tracking Regret: \\Theory, Algorithm, and  Applications to  Multi-Robot Coordination
}

\author{Zirui Xu, Hongyu Zhou, Vasileios Tzoumas
\thanks{Manuscript received August 31, 2022; Revised December 5, 2022; Accepted February 1, 2023. This paper was recommended for publication by Editor M. Ani Hsieh upon evaluation of the Associate Editor and Reviewers’ comments.}
\thanks{The authors are with the Department of Aerospace Engineering, University of Michigan, Ann Arbor, MI 48109 USA;  {\tt\footnotesize \{ziruixu,zhouhy,vtzoumas\}@umich.edu}}
\thanks{Digital Object Identifier (DOI): see top of this page.}
}

%% file: preamble_symbols.tex
\newtheorem{theorem}{Theorem}
\newtheorem{problem}{Problem}

\newtheorem{definition}{Definition}
\newtheorem{proposition}{Proposition}

\newtheorem{remark}{Remark}

\newcommand{\bdmath}{\begin{dmath}}
\newcommand{\edmath}{\end{dmath}}
\newcommand{\beq}{\begin{equation}}
\newcommand{\eeq}{\end{equation}}
\newcommand{\bdm}{\begin{displaymath}}
\newcommand{\edm}{\end{displaymath}}
\newcommand{\bea}{\begin{eqnarray}}
\newcommand{\eea}{\end{eqnarray}}
\newcommand{\beal}{\beq \begin{array}{lll}}
\newcommand{\eeal}{\end{array} \eeq}
\newcommand{\beas}{\begin{eqnarray*}}
\newcommand{\eeas}{\end{eqnarray*}}
\newcommand{\ba}{\begin{array}}
\newcommand{\ea}{\end{array}}
\newcommand{\bit}{\begin{itemize}}
\newcommand{\eit}{\end{itemize}}
\newcommand{\ben}{\begin{enumerate}}
\newcommand{\een}{\end{enumerate}}

\newcommand{\calA}{{\cal A}}
\newcommand{\calB}{{\cal B}}

\newcommand{\calN}{{\cal N}}

\newcommand{\calS}{{\cal S}}
\newcommand{\calT}{{\cal T}}

\newcommand{\calV}{{\cal V}}

\definecolor{myblue}{RGB}{65 105 225}

\newcommand{\hide}[1]{}

\newcommand{\hiddenText}{{\color{gray} hidden text.}}
\newcommand{\hideWithText}[1]{\hiddenText}

\newcommand{\opt}{^{\star}}

\newcommand{\scenario}[1]{{\smaller \sf#1}\xspace}



%% file: shortcuts.tex
\newcommand{\ie}{\emph{i.e.},\xspace}
\newcommand{\eg}{\emph{e.g.},\xspace}
\newcommand{\myin}{\, \in \,}

\newcommand{\sg}{\scenario{SG}}

\newcommand{\sgd}{{\smaller\scenario{\widehat{SG}}}}

\newcommand{\myParagraph}[1]{{\bf #1.}\xspace}

\renewcommand{\opt}{\scenario{OPT}}

\newcommand{\fsf}[1]{\scenario{FSF$^{\star}|_{#1}$}}
\newcommand{\solosg}{\calA^{\onalg}}
\newcommand{\distfsf}{p}
\newcommand{\onalg}{\scenario{OSG}}
\newcommand{\solopt}{\calA^{\opt}}

%% file: Abstract.tex
\begin{abstract}
We enable efficient and effective coordination in unpredictable environments, \ie in environments whose future evolution is unknown~a priori and even adversarial.  
We are motivated by the future of autonomy that involves multiple robots coordinating in dynamic, unstructured, and adversarial environments to complete  complex tasks such as target tracking, {environmental mapping}, and area monitoring.  Such tasks are often modeled as submodular maximization coordination problems.
We introduce the first submodular coordination algorithm with bounded \textit{tracking regret}, \ie with bounded suboptimality with respect to optimal time-varying actions that know the future a priori.  
The bound gracefully degrades with the environments' capacity to change adversarially.  It also quantifies how often the robots must re-select actions to ``learn'' to coordinate as if they knew the future a priori. The algorithm requires the robots to select actions sequentially based on the actions selected by the previous robots in the sequence.   Particularly, the algorithm generalizes the seminal {Sequential Greedy} algorithm by Fisher et al.~to unpredictable environments, leveraging submodularity and algorithms for the problem of \textit{tracking the best expert}.
We validate our algorithm in simulated scenarios of target tracking.
\end{abstract}

\vspace{-2mm}

%% file: 1-Introduction.tex
\vspace{-6mm}
\section{Introduction}\label{sec:Intro}
\IEEEPARstart{I}{n} the future, robots will be jointly planning actions to complete complex tasks such as:
\begin{itemize}[leftmargin=9pt]
\item \textit{Target Tracking}: How mobile robot networks can collaboratively track multiple evading targets?~{\cite{corah2021scalable}}
    \item \textit{{Environmental Mapping}}: How mobile robots can collaboratively map an unknown environment?~\cite{atanasov2015decentralized}
    \item \textit{Area Monitoring}: How robot swarms can collaboratively monitor an area of interest?~\cite{schlotfeldt2021resilient}
\end{itemize}
 
All the aforementioned coordination tasks have been modeled by researchers in robotics, control, and machine learning  via optimization problems of the form
\begin{equation}\label{eq:intro}
\max_{a_{i,\,t}\,\in\,\mathcal{V}_i,\,  \forall\, i\,\in\, \calN}\
         f_t(\,\{a_{i,\,t}\}_{i\myin \calN}\,), \ \ t=1,2,\ldots,
\end{equation}
where $\calN$ is the robot set, $a_{i,\,t}$ is robot $i$'s action at time step $t$, $\calV_i$ is robot $i$'s set of available actions, 
and $f_t:2^{\prod_{i \in \calN}\calV_i}\mapsto\mathbb{R}$ is the objective function that captures the task utility.  Particularly, $f_t$ is considered computable prior to each time step~$t$ given a model about the future evolution of the environment~\cite{krause2008near,singh2009efficient,tokekar2014multi,atanasov2015decentralized,gharesifard2017distributed,grimsman2018impact,corah2018distributed,corah2019distributed,zhou2018resilient,corah2021scalable,schlotfeldt2021resilient}; \eg in target tracking, a stochastic model for the targets' future motion is often considered available, and then $f_t$ can be chosen for example as the mutual information between the position of the robots and that of the targets~\cite{atanasov2015decentralized}. 

Although \cref{eq:intro} is generally NP-hard~\cite{Feige:1998:TLN:285055.285059}, near-optimal polynomial-time approximation algorithms have been proposed when $f_t$ is \textit{submodular}~\cite{fisher1978analysis}, a diminishing returns property.  For example, the \textit{Sequential Greedy} (\sg) algorithm~\cite{fisher1978analysis} achieves the near-optimal $1/2$ approximation bound when $f_t$ is submodular.  All aforementioned complex tasks can be modeled as submodular coordination problems, and thus \sg and its variants are commonly used in the literature~\cite{krause2008near,singh2009efficient,tokekar2014multi,atanasov2015decentralized,gharesifard2017distributed,grimsman2018impact,corah2018distributed,corah2019distributed,zhou2018resilient,corah2021scalable,schlotfeldt2021resilient}.

But complex tasks often evolve in environments that change unpredictably, \ie in environments whose future evolution is unknown a priori.  For example, during adversarial target tracking the targets’ actions can be unpredictable since their intentions and maneuvering capacity may be unknown~\cite{sun2020gaussian}. In such challenging environments, the robots that are tasked to track the targets cannot simulate the future to compute $f_t$ prior to time step $t$. Hence, the robots have to coordinate their actions by relying on past information only, \eg by relying only on the retrospective utility of their actions once the evolution of the environment has been observed.

In this paper, we aim to solve \cref{eq:intro} in unpredictable environments where 
 $f_t$ is unknown to the robots prior to time step $t$ and thus the robots need to coordinate actions by relying only on the retrospective utility of their actions.  Our goal is to provide polynomial-time algorithms with bounded suboptimality with respect to optimal time-varying multi-robot actions that know the future a priori, \ie with bounded \emph{tracking regret}~\cite{cesa2006prediction} ---the optimal actions ought to be time-varying to be effective against a changing environment such as an evading target.  To this end, we will leverage tools from the literature of online learning~\cite{cesa2006prediction}.

\myParagraph{Related Work} 
The current algorithms for \cref{eq:intro}
either (i) assume that $f_t$ is known prior to each time step $t$, instead of unknown, or (ii) apply to static environments where the optimal actions are time-invariant, instead of time-varying, or (iii) run in exponential time, instead of polynomial. 
No polynomial-time coordination algorithm exists that addresses \cref{eq:intro} when $f_t, t=1,2,\ldots$, are unknown a priori and where the optimal actions in hindsight are time-varying:

\paragraph{Related Work in {Submodular Optimization for Multi-Robot Coordination}} {The seminal algorithm \textit{Sequential Greedy} (\sg)~\cite{fisher1978analysis} is the first polynomial-time algorithm for \cref{eq:intro} with near-optimal approximation guarantees. Algorithms based on \sg have enabled} multi-robot coordination for a spectrum of tasks from target tracking \cite{tokekar2014multi,zhou2018resilient,corah2021scalable} and environmental exploration~\cite{corah2019distributed,liu2021distributed} to collaborative mapping \cite{atanasov2015decentralized, schlotfeldt2021resilient} and area monitoring \cite{gharesifard2017distributed,corah2018distributed,robey2021optimal,rezazadeh2021distributed,konda2022execution,xu2022resource}. However, these algorithms assume {a priori \textit{known} $f_t$.}

\paragraph{Related Work in {Online Learning for Submodular Optimization with Static Regret}}   Online learning algorithms have been proposed for \cref{eq:intro} to account for the case where $f_t$ is unknown a priori~\cite{streeter2008online,streeter2009online,suehiro2012online,golovin2014online,chen2018online,zhang2019online}. But these algorithms apply only to tasks where the optimal solution is \textit{static}: they guarantee bounded suboptimality with respect to optimal time-\underline{in}variant robot actions, instead of  optimal time-varying ones.

\paragraph{Related Work in {Online Learning of Time-Varying Optimal Actions}} The problem of learning online a sequence of time-varying actions that are optimal in hindsight constitutes the problem of \textit{tracking the best expert}~\cite{herbster1998tracking,gyorgy2005tracking,gyorgy2007line,zhang2017improved,zhang2018adaptive,harvey2020improved,matsuoka2021tracking}.
The problem involves an agent that selects actions online to maximize an accumulated utility across a number of time steps. {The challenge is that} the utility associated with each action is unknown a priori.
Although algorithms for the \textit{tracking the best expert} problem can be applied ``as is'' to \cref{eq:intro}, they then require {exponential time} to run, instead of polynomial.
Similarly, although \cite{matsuoka2021tracking} recently leveraged such algorithms to provide polynomial-time online learning algorithms for the problem of \textit{cardinality-constrained submodular maximization}, since this problem takes the form of $\max_{\calS\,\subseteq\, \calV, \, |\calS|\,\leq\, k}\, f(\calS)$ given an integer $k$ and a function $f:2^\calV\mapsto \mathbb{R}$, those algorithms in~\cite{matsuoka2021tracking} cannot be applied to \cref{eq:intro}, which has the form $\max_{a_{i,\,t}\,\in\,\mathcal{V}_i,\,  \forall\, i\,\in\, \calN}\;f(\,\{a_{i,\,t}\}_{i\myin \calN}\,)$, Instead, inspired by~\cite{matsuoka2021tracking}, our algorithm addresses the latter problem.

\myParagraph{Contribution}
We provide the first polynomial-time online learning algorithm with bounded tracking regret for multi-robot submodular coordination in unpredictable environments
(\Cref{sec:algorithm}).  We name the algorithm \textit{Online Sequential Greedy} (\onalg).  The algorithm generalizes the  \textit{Sequential Greedy} algorithm~\cite{fisher1978analysis} from
the setting where each $f_t$ is known a priori to the online setting
where $f_t$ is unknown a priori.
{As such, the algorithm requires the robots to select actions sequentially based on the actions selected by the previous robots in the sequence.} 
\onalg enjoys the properties: 

\begin{itemize}[leftmargin=9pt]
    \item \textit{Efficiency}: {For each agent $i$, \onalg  has a running time linear in the number of available actions per time step} (\Cref{subsec:complexity}).

    \item \textit{Effectiveness}: \onalg guarantees  bounded tracking regret (\Cref{subsec:tracking-regret}). The bound gracefully degrades with the environments' capacity to change adversarially.  It quantifies the intuition that the agents should be able to effectively adapt to an unpredictable environment when they {can} re-select actions frequently enough with respect to the environment's rate of change.  Specifically, the bound guarantees asymptotically and in expectation that the agents select actions near-optimally as if they knew the future a priori, matching the performance of the Sequential Greedy algorithm~\cite{fisher1978analysis}. 
\end{itemize}

{Inspired by~\cite{matsuoka2021tracking},} our technical approach innovates by leveraging algorithms for the  \textit{tracking the best expert} problem~\cite{matsuoka2021tracking}, and the submodularity of the objective functions. Although the direct application of the \textit{tracking the best expert} framework to \cref{eq:intro} results in an exponential-time algorithm, by leveraging submodularity, we obtain the linear-time \onalg.  

\myParagraph{Numerical Evaluations} We evaluate \onalg in simulated scenarios of two mobile robots pursuing two mobile targets (\Cref{sec:experiments}).  We consider non-adversarial and adversarial targets:  the non-adversarial targets traverse predefined trajectories, independently of the robots' motion; whereas, the adversarial targets maneuver, in response to the robots' motion.  In both cases, the targets' future motion and maneuvering capacity are unknown to the robots.
Across the simulated scenarios, \onalg enables the robots to closely track the targets despite being oblivious to the targets' future motion. 

%% file: 2-Problem.tex
\section{Online Submodular Coordination with Bounded Tracking-Regret}\label{sec:problem}

We define the problem \textit{Online Submodular Coordination with Bounded Tracking-Regret}. To this end, we set:
\begin{itemize}
    \item $\calV_\calN \triangleq \prod_{i\myin \calN} \,\calV_i$ is the set of possible action combinations for all the agents $\calN$, given the set of available actions $\calV_i$ for each agent $i\in\calN$;
    \item $f(\,a\;|\;\calA\,)\triangleq f(\,\calA \cup \{a\}\,)-f(\,\calA\,)$ is the marginal gain of adding $a$ to $\calA$, given an objective set function $f:2^\calV\mapsto \mathbb{R}$,  $a \in \calV$, and $\calA \subseteq\calV$.
    \item $|\calA|$ is the cardinality of $\calA$, given a discrete set $\calA$. 
\end{itemize}

The following framework is also considered.

\myParagraph{Agents}  $\calN$ is the set of all agents. The terms ``\emph{agent}'' and ``\emph{robot}'' are used interchangeably in this paper.   
The agents coordinate actions via a coordinate descent scheme commonly used in the literature~\cite{singh2009efficient,tokekar2014multi,atanasov2015decentralized,gharesifard2017distributed,grimsman2018impact,corah2018distributed,corah2019distributed,zhou2018resilient,corah2021scalable,schlotfeldt2021resilient} where the agents sequentially choose actions based on the actions selected by all previous agents in the sequence.

\myParagraph{Actions} $\calV_i$ is a \textit{discrete} and \textit{finite} set of actions available to robot $i$.  For example, $\calV_i$ may be a set of (i) motion primitives  that robot $i$ can execute to move in the environment~\cite{tokekar2014multi} or (ii) robot $i$'s discretized control inputs~\cite{atanasov2015decentralized}.

\myParagraph{Objective Function} The robots coordinate their actions to maximize an objective function.  In information-gathering
tasks such as target tracking, {environmental mapping}, and area monitoring, typical objective functions are the \emph{covering functions} \cite{corah2018distributed,robey2021optimal, downie2022submodular}.  Intuitively, these functions capture how much area{/information} is covered given the actions of all robots.  
They satisfy the properties defined below (\Cref{def:submodular}).

\begin{definition}[Normalized and Non-Decreasing Submodular Set Function{~\cite{fisher1978analysis}}]\label{def:submodular}
A set function $f:2^\calV\mapsto \mathbb{R}$ is \emph{normalized and non-decreasing submodular} if and only if 
\begin{itemize}
\item $f(\,\emptyset\,)=0$;
\item $f(\,\calA\,)\leq f(\,\calB\,)$, for any $\calA\subseteq \calB\subseteq \calV$;
\item $f(\,s\;|\;\calA\,)\geq f(\,s\;|\;{\mathcal{B}}\,)$, for any $\calA\subseteq {\mathcal{B}}\subseteq\calV$ and $s\in \calV$.
\end{itemize}
\end{definition}

Normalization $(f(\,\emptyset\,)=0)$ holds without loss of generality.  In contrast, monotonicity and submodularity are intrinsic to the function.  
Intuitively, if $f(\,\calA\,)$ captures the area \emph{covered} by a set $\calA$ of activated cameras, then the more sensors are activated, the more area is covered; this is the non-decreasing property.  Also, the marginal gain of the covered area caused by activating a camera $s$ \emph{drops} when \emph{more} cameras are already activated; this is the submodularity~property.

\myParagraph{Problem Definition}  In this paper, we focus on:

\begin{problem}[Online Submodular Coordination]
\label{pr:online}
Assume a time horizon $H$ of operation discretized to $T$ time steps.  The agents select actions $\{a_{i,\;t}\}_{i\,\in\,\calN}$ \emph{online} such that at each time step $t=1,\ldots, T$ they solve the optimization problem
\begin{equation}
    \max_{a_{i,\,t}\,\in\,\mathcal{V}_i, \,
         \forall\, i\,\in\, \calN}\
    f_t(\,\{a_{i,\,t}\}_{i\myin \calN}\,),
\end{equation}
where $f_t: 2^{\calV_\calN}\mapsto \mathbb{R}$ is a normalized and non-decreasing submodular set function, becoming known to the agents only \emph{once} they have executed their actions $\{a_{i,\,t}\}_{i\myin \calN}$.
\end{problem}

{\Cref{pr:online} assumes that the agents know the full objective function $f_t: 2^{\prod_{i \in \calN}\calV_i}\mapsto\mathbb{R}$ once they have executed their actions for the time step $t$.  This assumption is known as the \emph{full information} setting in the literature of online learning~\cite{matsuoka2021tracking}.}

\begin{remark}[{Feasibility of the Full Information Setting}]\label{rem:full-info}
The full information setting is feasible in practice when the agents can simulate the past upon observation of the environments' new state by the end of any time step $t$.  
For example, during target tracking with multiple robots, once the robots have executed their actions and observed the targets’ new positions by the end of the time step $t$, then they can evaluate in hindsight the effect of all possible actions they could have selected instead. That is, $f_t$ becomes fully known after the robots have acted at time step $t$.
\end{remark}

\begin{remark}[Adversarial Environment]\label{rem:adversarial}
The objective function $f_t$ can be \emph{adversarial}, \ie the environment may choose $f_t$ once it has observed the agents' actions $\{a_{i,\,t}\}_{i\myin \calN}$.  When $f_t$ changes arbitrarily bad between time steps, then inevitably no algorithm can guarantee near-optimal performance.  In this paper, we provide a randomized algorithm \emph{(\onalg)} which guarantees in expectation a suboptimality bound that deteriorates gracefully as the environment becomes more adversarial.
\end{remark}

%% file: 3-Algorithms.tex
\section{Online Sequential Greedy (\onalg) Algorithm} \label{sec:algorithm}

We present Online Sequential Greedy (\onalg). 
\onalg leverages as subroutine an algorithm for the problem of {tracking the best expert}~\cite{matsuoka2021tracking}.  
Thus, before presenting \onalg in \Cref{subsec:osg}, we first present the \textit{tracking the best expert} problem in \Cref{subsec:tracking-best-expert}, along with its solution algorithm.

\subsection{The Problem of {Tracking the Best Expert}}\label{subsec:tracking-best-expert}

\input{alg-fsfstar}

\textit{Tracking the best expert} involves an agent selecting actions to maximize a total reward (utility) across a given number of time steps.  The challenge is that the reward associated with each action is time-varying and unknown to the agent before the action has been executed. Therefore, to solve the problem, the agent needs to somehow guess the  ``best expert'' actions, \ie the actions achieving the highest reward at each time step.

To formally state the problem, we use the notation:
\begin{itemize}
    \item $\calV$ denotes the set of actions available to the agent;
    \item $r_{i,\,t}$ denotes the reward that the agent receives by selecting action $i\in\calV$ at the time step $t$;
    \item $r_t\triangleq \left[\,r_{1,\,t}, \dots, r_{|\calV|,\,t}\,\right]^\top$ is the vector of all rewards at $t$;
    \item $i^\star_t\in\arg\max_{i\myin\calV}\;r_{i,\,t}$; \ie $i^\star_t$ is the index of the ``best expert'' action at time $t$, that is, of the action that achieves the highest reward at time $t$;
    \item ${\bf 1}(\cdot)$ is the indicator function, \ie ${\bf 1}(x)=1$ if the event $x$ is true, otherwise ${\bf 1}(x)=0$.
    \item $P(T) \triangleq \sum_{t=1}^{T-1} {\bf 1}(\,i^\star_t\neq i^\star_{t+1}\,)$; \ie $P(T)$ counts how many times the best action changes over $T$ time steps.
\end{itemize}

\begin{problem}[Tracking the Best Expert \cite{cesa2006prediction}]\label{pr:tracking-the-best-expert}
Assume a time horizon $H$ of operation discretized to $T$ time steps. The agent selects an action $a_{t}$ \emph{online} at each time step $t=1,\ldots, T$ to solve the optimization problem
\vspace{-1mm}
\begin{equation}
    \max_{a_{t}\myin\mathcal{V},\, t\,=\,1,\ldots,T} \;\; \sum_{t=1}^T\;r_{a_t,\,t},
    \vspace{-1mm}
\end{equation}
where the rewards $\{r_{i,\,t}\}_{i\,\in\,\calV}$ can take any value and become known to the agent only \emph{once} the agent has executed action~$a_{t}$.
\end{problem} 

\Cref{alg:FSFstar}, {whose intuitive description is offered at the end of this subsection,} guarantees a suboptimality bound when solving \Cref{pr:tracking-the-best-expert} despite the challenge that the ``best expert'' actions $i^\star_1, \ldots, i^\star_T$ are unknown a priori to the agent, \ie despite the challenge that the rewards $\{r_{i,\,t}\}_{i\,\in\,\calV}$ become known only \emph{once} the agent has executed action $a_{t}$. To this end, the algorithm provides the agent with a probability distribution $p_t$ over the action set $\calV$ at each time step $t$, from which the agent draws an action $a_t$.
Then, in expectation the agent's total reward is guaranteed to be~\cite[Corollary 1]{matsuoka2021tracking}: 
\begin{equation}\label{eq:FSFstar_bound}
     \sum_{t=1}^{T} \;r_{t}^{\top}\,p_t \;\geq\; \sum_{t=1}^{T} \;r_{i^\star_t,\, t} - \tilde{O}\left[\,\sqrt{T\,P(T)}\,\right],
\end{equation}
where $\tilde{O}\left[\cdot\right]$ hides $\log$~terms. Thus, if $i^\star_t$ does not change many times across consecutive time steps $t$, particularly, if $P(T)$ grows slow enough with $T$, then in expectation the agent is able to track the ``best expert'' actions as $T$ increases: if $\sqrt{T\,P(T)}/T\rightarrow 0$ as $T\rightarrow +\infty$, then \cref{eq:FSFstar_bound} implies that $r_t^\top\,p_t\rightarrow r_{i^\star_t,\,t}$, \ie in expectation $a_t\rightarrow i^\star_t$.

\begin{remark}[Randomization Need]\label{rem:randomization}
\Cref{alg:FSFstar} is randomized since $i^\star_1, i^\star_2, \ldots$ may evolve adversarially, \ie may adapt to the agent's selected actions such that to minimize the agent's total reward. If an algorithm for \Cref{pr:tracking-the-best-expert} is instead deterministic, then the environment can minimize the agent's total reward, making a bound similar to \cref{eq:FSFstar_bound} impossible. For example, if a deterministic algorithm picks actions for the time step $t$ based on the observed rewards at time step $t-1$, then the environment can adapt the rewards $\{r_{i,\,t}\}_{i\,\in\,\calV}$ at time $t$ such that $r_{i^\star_{t-1},\,t}$ is the minimum reward among all $\{r_{i,\,t}\}_{i\,\in\,\calV}$.
\end{remark}

{\myParagraph{Intuition Behind \Cref{alg:FSFstar}}} \Cref{alg:FSFstar} computes the probability distribution $p_t$ online given the  observed rewards, \ie given $\{r_{i,\,\cdot}\}_{i\,\in\,\calV}$ up to $t-1$ (lines 5-15). To this end, \Cref{alg:FSFstar} assigns each action $i \in \calV$  time-dependent weights $\{w_{i,\,t}^{(j)}\}_{j\,=\,1,\ldots,J}$ that increase the higher $r_{i,\,1},\ldots, r_{i,\,t-1}$ are compared with the corresponding rewards of all other actions (lines 11-14).  The effect that the older rewards have on the value of $w_{i,\,t}^{(j)}$ is controlled by the parameter $\gamma^{(j)}$ (lines 11-14). $\gamma^{(j)}$  can be interpreted as a ``learning'' rate: the higher the $\gamma^{(j)}$ is, the more $w_{i,\,t}^{(j)}$ depends on the most recent rewards only, causing $w_{i,\,t}^{(j)}$ to adapt to (``learn'') the recent environment faster.  Thus, higher values of $\gamma^{(\cdot)}$ are desirable when the environment is more adversarial, \ie  when $P(T)$ is larger.
To account for that $P(T)$ is unknown a priori, \Cref{alg:FSFstar} computes in parallel multiple weights, the  $\{w_{i,\,t}^{(j)}\}_{j\,=\,1,\ldots,J}$, corresponding to the multiple learning rates $\{\gamma^{(j)}\}_{j\,=\,1,\ldots,J}$. %
$\{\gamma^{(j)}\}_{j\,=\,1,\ldots,J}$ cover the spectrum from small to large sufficiently enough (lines 1-2) since \cref{eq:FSFstar_bound} achieves the
best known bound up to $\log$ factors~\cite{robinson2021improved}.

\subsection{The OSG Algorithm}\label{subsec:osg}

\input{alg-osg}

\onalg is presented in \Cref{alg:online}. 
\onalg generalizes the Sequential Greedy 
(\sg) algorithm~\cite{fisher1978analysis} to the online setting of \Cref{pr:online}, leveraging at the agent-level \scenario{FSF$^\star$} (\Cref{alg:FSFstar}).\footnote{Although \Cref{pr:online} is a special case of \Cref{pr:tracking-the-best-expert}, the direct application of \Cref{alg:FSFstar} to \Cref{pr:online} results in an exponential time algorithm (since then the single agent in \Cref{pr:tracking-the-best-expert} is replaced by the multi-robot system in \Cref{pr:online} whose number of available actions at each time step is $\prod_{i\,\in\,\calN}\;|\calV_i|$). \onalg instead has linear computational complexity (\Cref{pror:computations}).}   
Particularly, when $f_t$ is known a priori, instead of unknown per \Cref{pr:online}, then \sg
instructs the agents to sequentially select actions $\{a_{i,\,t}^\sg\}_{i\,\in\,\calN}$ at each $t$ such that 
\begin{equation}\label{eq:sga}
  a_{i,\,t}^\sg \,\in\, \max_{a\myin\calV_i}\;\; f_t(\,a\;|\;\{a_{1,\,t}^\sg,\ldots,a_{i-1,\,t}^\sg\}\,),
\end{equation}
\ie agent $i$ selects $a_{i,\,t}^\sg$ after agent $i-1$, {given the actions of all previous agents  $\{1,\ldots, i-1\}$,} and  such that $a_{i,\,t}^\sg$ maximizes the marginal gain given the actions of all previous agents from $1$ through $i-1$. But since $f_t$ is unknown and adversarial per \Cref{pr:online}, \onalg  replaces the deterministic action-selection rule of \cref{eq:sga} with a \textit{tracking the best expert} rule (cf.~\Cref{rem:randomization}).  {Thus, \onalg is also a sequential action-selection algorithm}.

In more detail, \onalg starts by instructing each agent $i\in \calN$ to initialize an \scenario{FSF$^\star$} ---we denote the \scenario{FSF$^\star$} onboard each agent $i$ by \fsf{i}.  Specifically, agent $i$ initializes \fsf{i} with its action set $\calV_i$ and with the number $T$ of total time steps (line 1).  Then, at each time step $t=1,\ldots, T$, in sequence:
\begin{itemize}
    \item Each agent $i$ draws an action $a_{i,\, t}^\onalg$ given the  probability distribution $\distfsf_t^{(i)}$ output by \fsf{i}(lines 5-8).
    \item {All agents execute their actions $\{a_{i,\, t}^\onalg\}_{i\,\in\,\calN}$ and then observe $f_t: 2^{\calV_\calN}\mapsto \mathbb{R}$ (lines 9-10).}
    \item {Each agent $i$ receives from agent $i-1$ the actions of all agents with a lower index, $\solosg_{i-1,\,t}$,}
and then computes the marginal gain $f_t(\,a\,|\,\solosg_{i-1,\,t}\,)$ of each of its actions $a\in\calV_i$ (lines 11-16). Thus, in this step, each agent $i$ imitates in hindsight \sg's rule in \cref{eq:sga}.
    \item  Finally, each agent $i$ makes the vector of its computed marginal gains observable to \fsf{i} per the line 8 of \Cref{alg:FSFstar} (lines 17-18).  With this input, \fsf{i} will compute $\distfsf_{t+1}^{(i)}$, \ie the probability distribution over the agent $i$'s actions for the next time step $t+1$.
\end{itemize}

%% file: alg-fsfstar.tex
\setlength{\textfloatsep}{3mm}
\begin{algorithm}[t]
	\caption{\mbox{Fixed Share Forecaster$^\star$ (\scenario{FSF$^\star$})}~\cite{matsuoka2021tracking}.
	}\label{alg:FSFstar}
	\begin{algorithmic}[1]
		\REQUIRE Time steps $T$; and action set $\calV$.
		\ENSURE Probability distribution $p_t \myin \{[0,1]^{|\calV|}:\|p_t \|_1=$ $1\}$ over the actions in $\calV$ at each $t=1,\ldots,T$.
		\medskip
		
		\STATE $J\gets\lceil\log_2{(\,T\,)}\rceil$, $\gamma\gets\sqrt{{\log_{e}{(\,J\,)}}\,/\,{T}}$, $\beta\gets{1}\,/\,{T}$;
		        \vspace{0.7mm}
		\STATE  $\gamma^{(j)}\gets\sqrt{{\log_{e}{(\,|\calV|\,T\,)}}\,/\,{2^{j-1}}}$ for all $j = 1, \dots, J$;
		\STATE Initialize $z_{1}=\left[z_{1,\,1},\dots,z_{J,\,1}\right]^\top$ with $z_{j,\,1}=1$,\\ for all $j = 1, \dots, J$;
		\STATE Initialize $w_{1}^{(j)}=\left[w_{1,\,1}^{(j)},\dots,w_{|\calV|,\,1}^{(j)}\right]^\top$ with $w_{i,\,1}^{(j)}=1$,\\ for all  $j = 1, \dots, J$ and $i = 1, \dots, |\calV|$;
        \FOR {each time step $t = 1, \dots, T$}
         \vspace{0.2mm}
        \STATE $q_t\gets{z_t}\,/\,{\|z_t\|_1}$, $p_t^{(j)}\gets{w_t^{(j)}}\,/\,{\|w_t^{(j)}\|_1}$,\\ for all $j = 1, \dots, J$;
          \vspace{0.2mm}
        \STATE $p_t \gets \sum_{j=1}^J\, q_{j,\,t}\,p_t^{(j)}$; 
        \STATE \textbf{observe} the rewards $\{r_{i,\,t}\}_{i\,\in\,\calV}$  that the agent recei- \\ves by selecting any action $i\in\calV$ at the step $t$; 
        \vspace{0.7mm}
        \STATE $r_{t}\gets \left[\,r_{1,\,t}, \dots, r_{|\calV|,\,t}\,\right]^\top$;
        \FOR {$j = 1, \dots, J$}
                \vspace{0.7mm}
        \STATE $v_{i,\,t}^{(j)}\gets w_{i,\,t}^{(j)}\,\exp{(\,\gamma^{(j)}\,r_{i,\,t}\,)}$, $i=1,\dots,|\calV|$;
        \STATE $W_t^{(j)}\gets v_{1,\,t}^{(j)}+\dots+v_{|\calV|,\,t}^{(j)}$;
        \STATE $w_{i,\,t+1}^{(j)}\!\gets\!\beta\,\frac{W_t^{(j)}}{|\calV|}+(1-\beta)\,v_{i,\,t}^{(j)}$, \!$i=1,\dots,|\calV|$;
        \STATE $z_{j,\,t+1}\gets z_{j,\,t}\exp{(\,\gamma \,r_t^\top \,p_t^{(j)}\,)}$;
		\ENDFOR
		\ENDFOR
	\end{algorithmic}
\end{algorithm}

%% file: alg-osg.tex
\setlength{\textfloatsep}{3mm}
\begin{algorithm}[t]
	\caption{\mbox{Online Sequential Greedy (\onalg).}
	}
	\begin{algorithmic}[1]
		\REQUIRE Time steps $T$; and agents' action sets $\{\mathcal{V}_i\}_{i \myin \calN}$.
		\ENSURE \!Agent actions $\{a_{i,\, t}^\onalg\}_{i\,\in\,\calN}$ at each $t=1,\ldots, T$.
		\medskip
		\STATE Each agent $i\in\calN$ initializes an \scenario{FSF$^\star$} with the value of the parameters $T$ and $\calV_i$.
		\STATE Denote the \scenario{FSF$^\star$} onboard agent $i$ by \fsf{i}.
		\STATE 
		{Order the agents in $\calN$ such that $\calN=\{1,\ldots, |\calN|\}$.}
		\FOR {\text{each time step} $t = 1, \dots, T$}
 			\FOR {$i = 1, \dots, |\calN|$}
				\STATE \textbf{get} the output $\distfsf_t^{(i)}$ from \fsf{i}; 
				\STATE \textbf{draw} an action $a_{i,\,t}^\onalg$ from the distribution $\distfsf_t^{(i)}$;
			\ENDFOR
			\STATE \textbf{execute} $\{a_{i,\, t}^\onalg\}_{i\,\in\,\calN}$;  
			\STATE \textbf{observe} the objective function $f_t: 2^{\calV_{\calN}}\mapsto \mathbb{R}$; 
			\STATE $\solosg_{0,\,t} \gets \emptyset$;
			\FOR {$i = 1, \dots, |\calN|$
			}
			\STATE $\solosg_{i,\,t} \gets \solosg_{i-1,\,t}\cup \{a_{i,\,t}^\onalg\}$;
   			\FOR {every $a \in \calV_i$}
			\STATE $r_{a,\,t}^{\hspace{.5pt}(i)}\gets f_t(\,a\;|\;\solosg_{i-1,\,t}\,)$; 
			\ENDFOR
			\STATE $r_{t}^{\hspace{.5pt}(i)}\gets \{r_{a,\,t}^{\hspace{.5pt}(i)}\}_{a\myin \calV_i}$;
			\STATE \textbf{input} $r_{t}^{\hspace{.5pt}(i)}$ to \fsf{i} (per line 8 of \Cref{alg:FSFstar});
			\ENDFOR
		\ENDFOR
	\end{algorithmic}\label{alg:online}
\end{algorithm}

%% file: 4-Guarantees.tex
\vspace{-3mm}
\section{Performance Guarantees of \onalg}\label{sec:guarantee}

We quantify \onalg's computational complexity and approximation performance  (\Cref{subsec:complexity,subsec:tracking-regret} respectively).  

\vspace{-1mm}
\subsection{Computational Complexity of \onalg}\label{subsec:complexity}

\onalg is the first algorithm for \Cref{pr:online} with polynomial computational complexity. 

\begin{proposition}[Computational Complexity]\label{pror:computations}
\onalg requires each agent $i \in \calN$ to perform $O(T\,|\calV_i|)$ function evaluations and $O(T\,\log{(T)})$ additions and multiplications. 
\end{proposition}

The proposition holds true since at each $t=1,\ldots,T$, \onalg requires each agent $i$ to perform $O(|\calV_i|)$ function evaluations to compute the marginal gains in \onalg's line 14 and $O(\log{(T)})$ additions and multiplications to run \fsf{i}.

\vspace{-1mm}
\subsection{Approximation Performance of \onalg}\label{subsec:tracking-regret}

We bound \onalg's suboptimality with respect to the optimal actions the agents' would select if they knew the $\{f_t\}_{t=1,\ldots,T}$ a priori. Particularly, we bound \onalg's \textit{tracking regret}, proving that it gracefully degrades with the environment's capacity to select $\{f_t\}_{t=1,\ldots,T}$ adversarially (\Cref{th:half}). 

To present \Cref{th:half}: first, we define tracking regret, {particularly, \textit{$1/2$-approximate tracking regret}} (\Cref{def:tracking_regret}); then, we quantify the environment's capacity to select $\{f_t\}_{t=1,\ldots,T}$ adversarially (\Cref{def:environment_change}).  To these ends, we use the notation:
\begin{itemize}
    \item $\solopt_t\in\arg\max_{a_{i,\,t}\,\in\,\mathcal{V}_{i},\, \forall\, i\,\in\, \calN} \;f_t(\,\{a_{i,\,t}\}_{i\,\in\, \calN}\,)$; \ie $\solopt_t$ is the optimal actions the agents' would select at the time step $t$ if they knew the $f_t$ a priori;
    \item $a_{i,\,t}^\opt$ is agent $i$'s action among the actions in $\solopt_t$;
    \item $\calA_t\triangleq \{a_{i,\,t}\}_{i\,\in\,\calN}$; \ie $\calA_t$ is the set of all agents' actions at time step $t$.
\end{itemize}

\begin{definition}[{$1/2$-Approximate Tracking Regret}]\label{def:tracking_regret}
Consider any sequence of action sets $\{\calA_t\}_{t\,=\,1,\ldots,T}$. Then, $\{\calA_t\}_{t\,=\,1,\ldots,T}$'s {$1/2$-approximate tracking regret} is\footnote{\Cref{def:tracking_regret} generalizes existing notions of tracking regret \cite{herbster1998tracking,matsuoka2021tracking} to the online submodular coordination \Cref{pr:online}.} 
\begin{align}\label{eq:regret}
    &{\scenario{Tracking}\text{-}\scenario{Regret}_T^{(1/2)}}(\,\{\calA_t\}_{t=1,\ldots,T}\,) \nonumber\\
    &\qquad\qquad\qquad\quad\triangleq \frac{1}{2}\sum_{t=1}^{T}\; f_t(\,\solopt_t\,)\, -\, \sum_{t=1}^{T}\;f_t(\,\calA_t\,).
\end{align}
\end{definition}

\Cref{eq:regret} evaluates $\{\calA_t\}_{t=1,\ldots,T}$'s suboptimality  with respect to the optimal actions $\{\calA_t^\opt\}_{t=1,\ldots,T}$ the agents' would select if they knew the $\{f_t\}_{t=1,\ldots,T}$ a priori. The optimal  value $\sum_{t=1}^{T} f_t(\,\solopt_t\,)$ is discounted by $1/2$ in \cref{eq:regret} since solving exactly \Cref{pr:online} is NP-hard even when $\{f_t\}_{t=1,\ldots,T}$ are known a priori~\cite{sviridenko2017optimal}.  {Specifically, the best possible approximation bound in polynomial time is the $1-1/e$~\cite{sviridenko2017optimal}, while the Sequential Greedy algorithm~\cite{fisher1978analysis}, which \onalg extends to the online setting, achieves the near-optimal bound $1/2$.  In this paper, we prove that \onalg can approximate \textit{Sequential Greedy}'s near-optimal performance by bounding \cref{eq:regret}.}

\begin{definition}[Environment's Total Adversarial Effect]\label{def:environment_change}
The environment's \emph{total adversarial effect} over $T$ time steps is\footnote{\Cref{def:environment_change} generalizes existing notions of an environment's \emph{total adversarial effect} \cite{matsuoka2021tracking} to the online submodular coordination \Cref{pr:online}.} 
\begin{equation}\label{eq:rate}
	\Delta(T) \triangleq \sum_{t=1}^{T-1} \sum_{i\myin \calN}\; {\bf 1}(\,a_{i,\,t}^\opt\neq a_{i,\;t+1}^\opt\,).
\end{equation}
\end{definition}

$\Delta(T)$ captures the environment's total effect in selecting $\{f_t\}_{t=1,\ldots,T}$ adversarially
by counting how many times the optimal actions of the agents must shift across the $T$ steps to adapt to the changing $f_{(\cdot)}$.  
That is, any changes that do not necessitate the agents’ to adapt are ignored.  

In sum, the larger is the environment capacity  to select $\{f_t\}_{t=1,\ldots,T}$ adversarially, the larger  $\Delta(T)$ is, and the more frequently the agents need to change actions to remain optimal.

\begin{theorem}[Approximation Performance]\label{th:half} \onalg instructs the agents to select actions, $\{a_{i,\,t}^\onalg\}_{i\myin\calN,\, t=1,\ldots,T}$  guaranteeing
\begin{align}
	&\mathbb{E}\left[\,{\scenario{Tracking}\text{-}\scenario{Regret}_T^{(1/2)}}(\,\{a_{i,\,t}^\onalg\}_{i\myin\calN,\, t=1,\ldots,T}\,)\,\right] \nonumber\\ &\qquad\qquad\qquad\quad\leq\tilde{O}\left\{\sqrt{\,|\calN|\,T\,\left[\,\Delta(T)+|\calN|\,\right]}\,\right\},\label{eq:th-bound}
\end{align}
where $\mathbb{E}\left[\,\cdot\,\right]$ denotes expectation with respect to \onalg's randomness, and $\tilde{O}\left\{\cdot\right\}$ hides $\log$~terms.
\end{theorem}

\Cref{th:half} bounds the expected tracking regret of \onalg. The bound is a function of the number of robots, the total time steps $T$, and the environment's total adversarial effect.  

The upper bound in \cref{eq:th-bound} implies that if the environment's total adversarial effect grows slow enough with $T$, then in expectation the agents are able to $1/2$-approximately track the unknown optimal actions $\calA_1^\opt,\ldots, \calA_T^\opt$: if  
\begin{equation}\label{eq:learning-rate}
\tilde{O}\left\{\sqrt{\,|\calN|\,T\,\left[\,\Delta(T)+|\calN|\,\right]}\,\right\}\Big/T\rightarrow 0, \text{ for $T\rightarrow +\infty$},
\end{equation}
then \cref{eq:th-bound} implies $f_t(\,\calA_t\,)\rightarrow 1/2 f_t(\,\calA_t^\opt\,)$ in expectation.\footnote{{Discretizing time finely, such that $T\rightarrow +\infty$ is feasible, is limited in practice by the robots' inability to replan and execute actions instantaneously.}}

For example, \cref{eq:learning-rate} holds true in environments whose evolution in real-time is unknown yet predefined, instead of being adaptive to the agents' actions.  Then, $\Delta(T)$ is uniformly bounded since increasing the discretization density of time horizon $H$, \ie increasing the number of time steps $T$, does not affect the environment's evolution.  Thus,
$\Delta(T)/T\rightarrow 0$ for $T\rightarrow +\infty$, which implies  \cref{eq:learning-rate}. The result agrees with the intuition that the agents should be able to adapt to an unknown but non-adversarial environment when they {re-select actions with high enough frequency.}

\begin{remark}[``Learning'' to be Near-Optimal] When \cref{eq:learning-rate} holds true, \onalg enables the agents to asymptotically ``learn'' to coordinate as if they knew $f_t, f_{t+1}, \ldots$ a priori, matching the performance of the near-optimal Sequential Greedy algorithm~\cite{fisher1978analysis}: the Sequential Greedy algorithm guarantees $f_t(\,\calA_t\,)\geq  1/2 f_t(\,\calA_t^\opt\,)$ when $f_t$ is known a priori, and \cref{eq:learning-rate} asymptotically guarantees $f_t(\,\calA_t\,)\gtrsim 1/2 f_t(\,\calA_t^\opt\,)$ in expectation despite 
$f_t$ is unknown a priori.
\end{remark}

%% file: 5-Experiments.tex
\section{Numerical Evaluation in Multi-Target Tracking Tasks with Multiple Robots}\label{sec:experiments}
\input{6-figure-target-tracking}

\input{6-figure-adversarial-target-tracking}

We evaluate \onalg in simulated scenarios of target tracking.  We first consider non-adversarial targets (\Cref{subsec:osg_tracking}), \ie targets whose motion is non-adaptive to the robots' motion. Then, we consider adversarial targets (\Cref{subsec:osg_sg_adversarial}), \ie targets whose motion adapts to the robots' motion.  

\vspace{1mm}
\myParagraph{Common Simulation Setup across Simulated Scenarios} We consider two robots pursuing two targets. Particularly: 

\paragraph{Robots} The robots can observe the exact location of the targets. The challenge is that the robots are unaware of the targets' future motion and, as a result, the robots cannot coordinate their actions by projecting where the targets are going to be.  Instead, the robots have to coordinate their actions based only on the history of past observations and somehow guess the future. 
To move in the environment, each robot $i\in\calN$ can perform either of the actions \{``upward'', ``downward'', ``left'', ``right''\} at a speed of \{$1$, $2$\} units/s. 

\paragraph{Targets} The targets are either non-adversarial or adversarial, moving on the same 2D plane as the robots.  The available actions to the targets in each of the two scenarios are described in \Cref{subsec:osg_tracking} and \Cref{subsec:osg_sg_adversarial}, respectively.

\paragraph{Objective Function} The robots coordinate their actions $\{a_{i,\,t}\}_{i\myin \calN}$ to maximize at each time step $t$\footnote{The objective function in \cref{eq:distance} is a non-decreasing and submodular function.  The proof is presented in Appendix B.}
\begin{equation}\label{eq:distance}
   {f_t(\,\{a_{i,\,t}\}_{i\myin \calN}\,)\,=\, \sum_{j\myin\calT}\;\max_{i\myin \calN} \; \frac{1}{d_t(a_{i,\,t},\,j)},}
\end{equation}
where {$d_t(a_{i,\,t},\,j)$ is the distance between robot $i$ and target $j$ after $a_{i,\,t}$ has been executed.} Hence, {$d_t(a_{i,\,t},\,j)$} is known to the robots only once the robots have executed their actions and the targets' locations at time step $t$ have been observed. 

By maximizing $f_t$, the robots aim to ``capture'' the targets: $f_t$ indicates that if robot $i$ keeps being the closest to target $j$, then the remaining robots receive no marginal gain by trying to approach target $j$ and thus they would want to approach other targets to maximize the objective. 

\myParagraph{Computer System Specifications} We performed all simulations in Python 3.9.7, on a MacBook Pro equipped with the Apple M1 Max chip and a 32 GB RAM. 

\textbf{Code.} Our code is available at: \href{https://gitlab.umich.edu/iral-code/ral23-online-submodular-coordination}{https://gitlab.umich.edu/iral-code/ral23-online-submodular-coordination}.

\subsection{Non-Adversarial Targets: Non-Adaptive Target Trajectories}\label{subsec:osg_tracking}

\myParagraph{Simulation Setup} 
{We consider two scenarios of non-adaptive target trajectories: 
(i) the targets traverse predefined straight lines over a time horizon $H=50$s  with speed $1$ unit/s (Fig.~\ref{fig:target_tracking}(a-c)), and (ii) the targets traverse  noisy rectangular-like trajectories over a time horizon $H=100$s (Fig.~\ref{fig:target_tracking}(g-i)); specifically, the rectangular-like trajectories are generated by targets that follow a nominal rectangular trajectory with speed $1$ unit/s while randomizing their lateral speed by sampling from a Gaussian distribution with zero mean and variance $2$.} 
For each case, we evaluate \onalg when the robots' action re-selection frequency varies from $10$Hz to $20$Hz to $50$Hz.

\myParagraph{Results} The simulation results are presented in Fig.~\ref{fig:target_tracking}.  They reflect the theoretical analyses in \Cref{subsec:tracking-regret}. At $10$Hz, the robots fail to ``learn'' the targets' future motion, failing to reduce their distance to them (Fig.~\ref{fig:target_tracking}(a,d,g,j)).  The situation improves at $20$Hz (Fig.~\ref{fig:target_tracking}(b,e,h,k)), and  even further at $50$Hz (Fig.~\ref{fig:target_tracking}(c,f,i,l)), in which case the robots closely track the targets. {The average minimum distances improve from $2$ ($10$Hz) to $1$ ($20$Hz) to $0.3$ ($50$Hz) unit for the line case (Fig.~\ref{fig:target_tracking}(d-f)), and from $8$ ($10$Hz) to $4$ ($20$Hz)  to $2$ ($50$Hz) units for the noisy-rectangular case (Fig.~\ref{fig:target_tracking}(j-l)).}

\subsection{Adversarial Targets: Adaptive Target Trajectories
}\label{subsec:osg_sg_adversarial}

\myParagraph{Simulation Setup} We consider 
targets that maneuver when the robots are close enough. 
As long as the robots are more than $1.5$ units away from a target $j$, the target $j$ will keep moving ``right'' on a nominal straight line at a speed $1$ unit/s.  But once a robot is within $1.5$ units away, then target $j$ will perform a maneuver: 
target $j$ will first choose from moving ``upward'' or ``downward'' at $2$ units/s for $1$s to maximize the distance from the robots, and then will move diagonally for $0.05$s to return back to the nominal path with vertical speed {$40$ units/s and horizontal speed to the ``right'' $30$ units/s.}
 
To demonstrate the need for randomization in adversarial environments (\Cref{rem:randomization}), we compare \onalg with a deterministic algorithm that selects actions at each  $t$ with respect to the previously observed $f_{t-1}$. We denote the algorithm by {\smaller$\sgd$} since it is a (heuristic) extension of the Sequential Greedy algorithm~\cite{fisher1978analysis} to \Cref{pr:online}'s setting per the rule:
\begin{equation}\label{eq:sgd}
  a_{i,\,t}^{{\sgd}} \,\in\, \max_{a\myin\calV_i}\;
  f_{t-1}(\,a\;|\;\{a_{1,\,t}^{\sgd},\ldots,a_{i-1,\,t}^{\sgd}\}\,).
\end{equation}
The Sequential Greedy algorithm~\cite{fisher1978analysis} instead selects actions per the rule in \cref{eq:sga}, given the a priori knowledge of $f_t$.

\myParagraph{Results} The simulation results are presented in Fig.~\ref{fig:adversarial}. 
\onalg performs better than {\smaller$\sgd$}, as expected since {\smaller$\sgd$} prescribes actions to the robots ``blindly'' by deterministically following the target's previous position, instead of accounting for the whole history of the targets' past motions as \onalg does: \onalg triggers $50\%$ more target maneuvers than {\smaller$\sgd$} ($31$ maneuvers in Fig.~\ref{fig:adversarial}(a) vs.~$20$ maneuvers in Fig.~\ref{fig:adversarial}(b)), implying \onalg tracks the targets closer than {\smaller$\sgd$}; 
particularly, once \onalg has ``learned'' the target’s future motion (after the $30$-th second in Fig.~\ref{fig:adversarial}(c)), {\onalg results in average minimum
distances to each target that are  $20$\% smaller than {\smaller$\sgd$}'s.}

%% file: 6-figure-target-tracking.tex
\begin{figure*}[t!]
    \captionsetup{font=footnotesize}
	\begin{center}
	\hspace{-9.1cm}\begin{minipage}{\columnwidth}
	\hspace{-2mm}		
 	\begin{minipage}{0.67\columnwidth}%
            \centering%
            \includegraphics[width=\columnwidth]{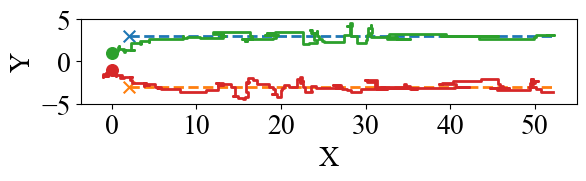} \\
            \vspace{-2mm}
            \caption*{(a) $H=50$s, $T=500$ time steps ($10$Hz).
            }
		\end{minipage}~
		\begin{minipage}{0.67\columnwidth}
            \centering%
            \includegraphics[width=\columnwidth]{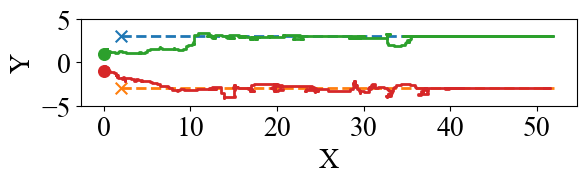} \\ 
            \vspace{-2mm}
            \caption*{(b) $H=50$s, $T=1000$ time steps  ($20$Hz). 
            }
		\end{minipage}~
		\begin{minipage}{0.67\columnwidth}%
            \centering%
            \includegraphics[width=\columnwidth]{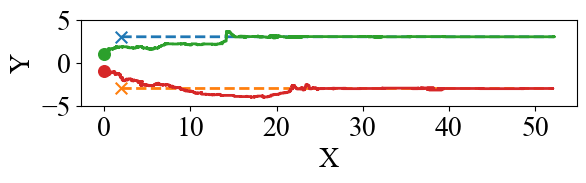} \\
            \vspace{-2mm}
            \caption*{(c) $H=50$s, $T=2500$ time steps ($50$Hz).
            }
		\end{minipage}
	\end{minipage} \\
	\vspace{1mm}
	\hspace{-8.9cm}\begin{minipage}{\columnwidth}
	\hspace{-2mm}		
 	\begin{minipage}{0.67\columnwidth}%
            \centering%
            \includegraphics[width=\columnwidth]{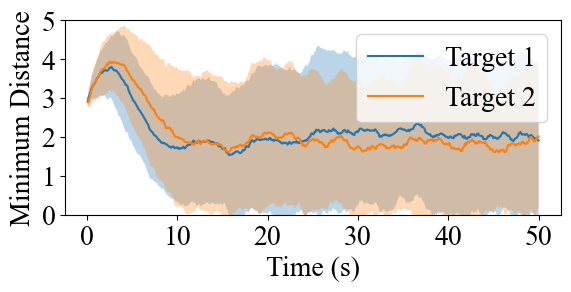} \\%
            \vspace{-2mm}
            \caption*{(d) $H=50$s, $T=500$ time steps ($10$Hz).
            }
		\end{minipage}~
		\begin{minipage}{0.67\columnwidth}%
		    \centering%
            \includegraphics[width=\columnwidth]{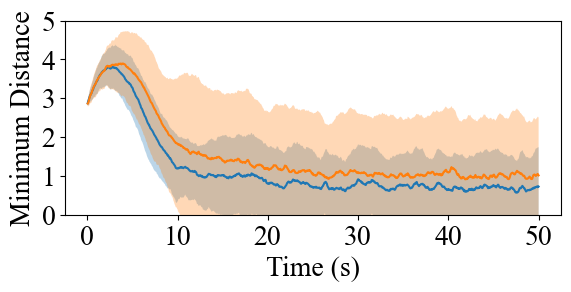} \\ 
            \vspace{-2mm}
            \caption*{(e) $H=50$s, $T=1000$ time steps ($20$Hz). 
            }
		\end{minipage}~
		\begin{minipage}{0.67\columnwidth}%
            \centering%
            \includegraphics[width=\columnwidth]{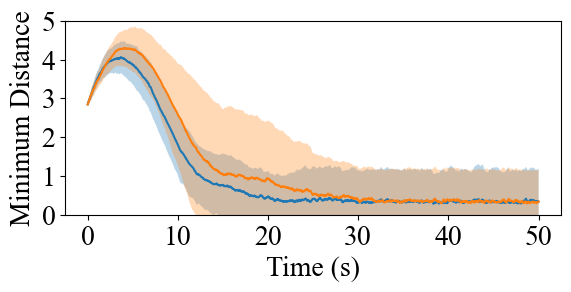} \\%
            \vspace{-2mm}
            \caption*{(f) $H=50$s, $T=2500$ time steps ($50$Hz).
            }
		\end{minipage}
	\end{minipage} \\
	\vspace{1mm}
	\hspace{-9cm}\begin{minipage}{\columnwidth}
	\hspace{-2mm}	
	\begin{minipage}{0.67\columnwidth}%
            \centering%
            \includegraphics[width=\columnwidth]{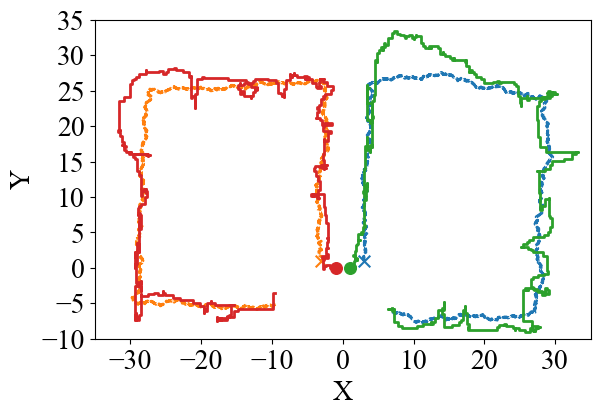} \\%
            \vspace{-2mm}
            \caption*{(g) $H=100$s, $T=1000$ time steps ($10$Hz).
            }
		\end{minipage}~
		\begin{minipage}{0.67\columnwidth}%
		    \centering%
            \includegraphics[width=\columnwidth]{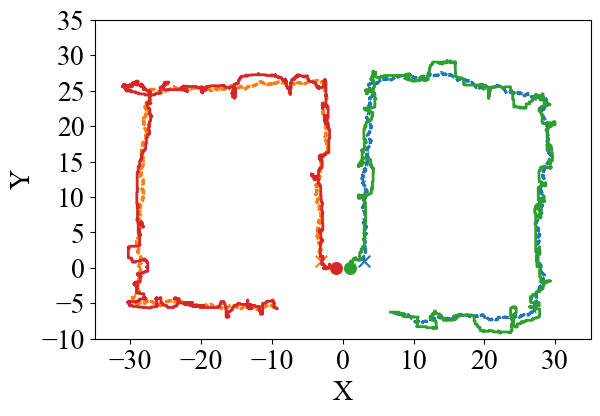} \\ 
            \vspace{-2mm}
            \caption*{(h) $H=100$s, $T=2000$ time steps ($20$Hz). 
            }
		\end{minipage}~
		\begin{minipage}{0.67\columnwidth}%
            \centering%
            \includegraphics[width=\columnwidth]{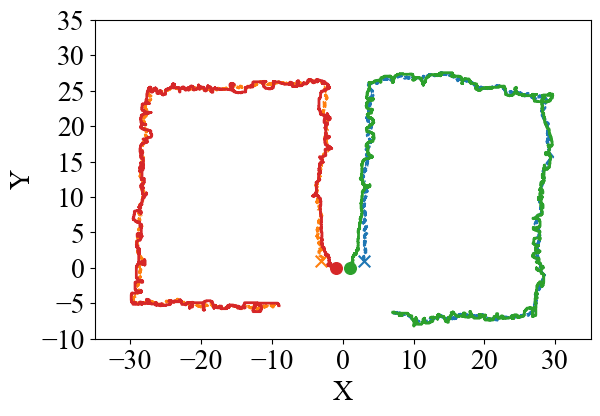} \\%
            \vspace{-2mm}
            \caption*{(i) $H=100$s, $T=5000$ time steps ($50$Hz).
            }
		\end{minipage}
	\end{minipage} \\
	\vspace{1mm}
	\hspace{-9.05cm}\begin{minipage}{\columnwidth}
	\hspace{-2mm}
	\begin{minipage}{0.67\columnwidth}%
            \centering%
            \includegraphics[width=\columnwidth]{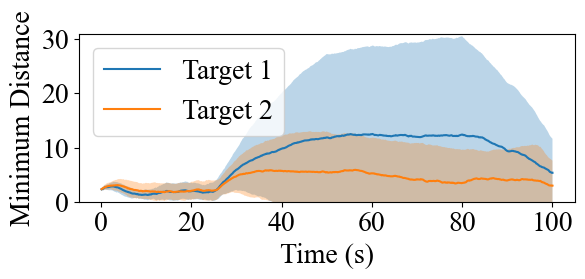} \\%
            \vspace{-2mm}
            \caption*{(j) $H=100$s, $T=1000$ time steps ($10$Hz).
            }
		\end{minipage}~
		\begin{minipage}{0.67\columnwidth}
		    \centering%
            \includegraphics[width=\columnwidth]{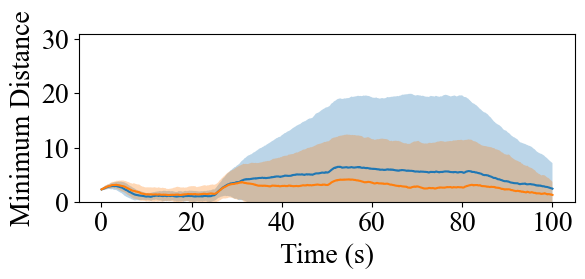} \\ 
            \vspace{-2mm}
            \caption*{(k) $H=100$s, $T=2000$ time steps ($20$Hz). 
            }
		\end{minipage}~
		\begin{minipage}{0.67\columnwidth}%
            \centering%
            \includegraphics[width=\columnwidth]{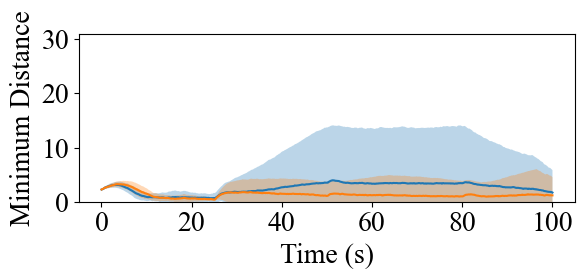} \\%
            \vspace{-2mm}
            \caption*{(l) $H=100$s, $T=5000$ time steps ($50$Hz).
            }
		\end{minipage}
	\end{minipage}
	\caption{\textbf{Non-Adversarial Target Tracking with Multiple Robots.} 2 robots pursue 2 targets that traverse trajectories that are non-adaptive to the robots motion: 
	(a)--(c) the targets traverse the dashed straight trajectories; (g)--(i) the targets traverse the dashed {noisy-rectangular trajectories that result from nominal rectangular trajectories corrupted with Gaussian noise of mean zero and variance $2$}.  The green and red solid lines are the robots' trajectories, starting from the solid dots, and the blue and orange dashed lines are the targets' trajectories, starting from the crosses. Across (a)--(c) and (g)--(i), the number $T$ of time steps per the given horizon $H$ varies, resulting in the robots to re-select actions with varying frequency. 
	(d)--(f) and (j)--(l) depict the minimum distance from each target to a robot, averaged over 50 instances of the simulation scenarios shown in (a)--(c) and (g)--(i) respectively.
	}\label{fig:target_tracking}
	\vspace{-7mm}
	\end{center}
\end{figure*}

%% file: 6-figure-adversarial-target-tracking.tex
\begin{figure*}[t]
    \captionsetup{font=footnotesize}
	\begin{center}
	\hspace{-9.1cm}\begin{minipage}{\columnwidth}		
	    \begin{minipage}{0.67\columnwidth}%
			\vspace{-.9mm}
            \centering%
            \includegraphics[width=\columnwidth]{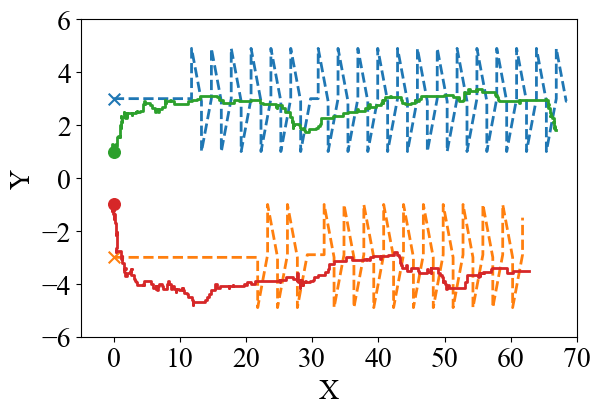} \\
            \vspace{-2mm}
            \caption*{(a) Target and Robot Trajectories per \onalg.
            }
		\end{minipage}~
		\begin{minipage}{0.67\columnwidth}%
			\vspace{-.9mm}
            \centering%
            \includegraphics[width=\columnwidth]{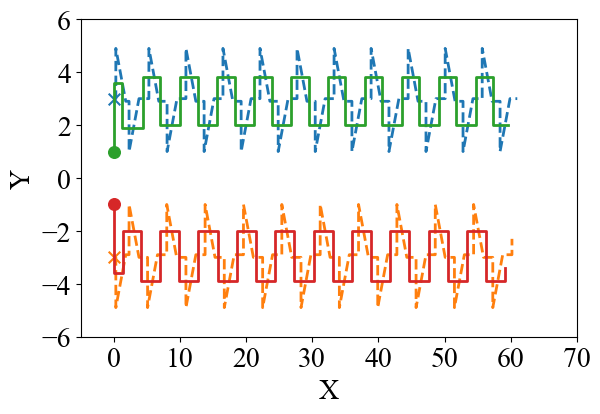} \\
            \vspace{-2.5mm}
            \caption*{(b) Target and Robot Trajectories per {\smaller$\sgd$}. 
            }
		\end{minipage}~
		\begin{minipage}{0.67\columnwidth}      
		\vspace{-.6mm}
		    \centering%
            \includegraphics[width=.97\columnwidth]{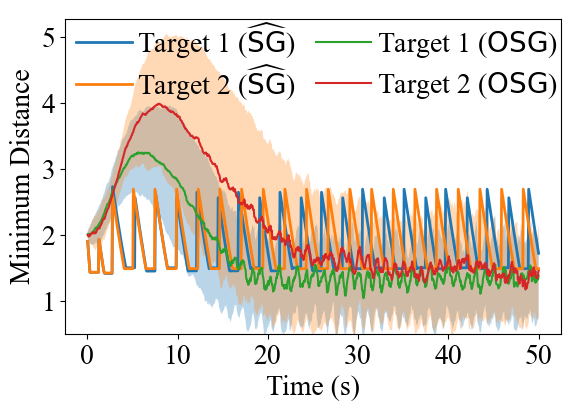} \\ 
            \vspace{-2mm}
            \caption*{(c) Minimum Distance.
            }
		\end{minipage}
	\end{minipage}
	\caption{\textbf{Adversarial Target Tracking with Multiple Robots.} 
	2 robots pursue 2 adversarial targets over a time horizon $H=50$s.  The robots re-select actions with frequency $20$Hz.  The targets maneuver up or down from a nominal straight line and then forward and back to the line every time a robot is at most 1.5 units away.  The robots select actions either per \onalg, inducing the trajectories in (a), or per {\smaller$\sgd$}, inducing the trajectories in (b).  The green and red solid lines are the robots' trajectories, starting from the solid dots, and the blue and orange dashed lines are the targets' trajectories, starting from the crosses.
    (c) depicts the minimum distance from each target to a robot, averaged over 50 instances of scenarios (a) and (b) respectively.  
	}\label{fig:adversarial}
	\vspace{-7mm}
	\end{center}
\end{figure*}

%% file: 6-Conclusion.tex
\section{Conclusion} \label{sec:con}
\myParagraph{Summary} 
We introduced the first algorithm for efficient and effective online submodular coordination in unpredictable environments (\onalg): \onalg is the first polynomial-time algorithm with bounded tracking regret for \Cref{pr:online}.
The bound gracefully degrades with the environments' capacity to change adversarially, quantifying how frequently the agents must re-select actions to ``learn'' to coordinate as if they knew the future a priori.  \onalg generalizes the seminal {Sequential Greedy} algorithm~\cite{fisher1978analysis} to \Cref{pr:online}'s online optimization setting. To this end, we leveraged the \scenario{FSF$^\star$} algorithm for the problem of \textit{tracking the best expert}.
We validated \mbox{\onalg in simulated scenarios of target tracking.}

\myParagraph{Future Work} We plan for the future work:

\paragraph{Partial Information Feedback} \onalg selects actions at each $t$ based on \textit{full information feedback}, \ie based on the availability of the functions $f_{k}:2^{\calV_\calN}\mapsto \mathbb{R}$ for each $k\leq t-1$.  The availability of these functions relies on the assumption that the agents can simulate the environment from the beginning of each step $k$ till its end.   
But the agents may lack the resources for this.  We will enable \onalg to rely only on \textit{partial information}, \ie only on the observed values $f_k(\,\{a_{i,\,k}\}_{i\,\in\,\calN}^\onalg\,)$ per the executed actions $\{a_{i,\,k}^\onalg\}_{i\,\in\,\calN}$.

\paragraph{Best of Both Worlds \emph{(BoBW)}} \onalg selects actions assuming the environment may evolve arbitrarily in the future.  The assumption is pessimistic when the environment's evolution is governed by a stochastic (yet unknown) model.  For example, \onalg's performance against the non-adversarial targets in \Cref{subsec:osg_tracking}, although it becomes near-optimal for large $T$, is pessimistic: the deterministic heuristic {\smaller$\sgd$} can be shown to perform better since by construction $f_{t-1}\simeq f_t$ in \Cref{subsec:osg_tracking}.  {We will extend \onalg such that it offers BoBW suboptimality guarantees~\cite{ito2021optimal}}.  BoBW guarantees become especially relevant under \textit{partial information feedback} since then heuristics such as the {\smaller$\sgd$} cannot apply in the first place.

%% file: new-App.tex
\section*{Appendix}\label{app:notation}
\subsection{Proof of \Cref{th:half}}

We use the notation:
\begin{itemize}
    \item $\solopt_{i-1,\,t}$ is the optimal solution set for the first $i-1$ agents at time step $t$;
    \item $\Delta_i(T) \triangleq \sum_{t=1}^{T-1} {\bf 1}(a_{i,\,t}^\opt\neq a_{i,\;t+1}^\opt)$ is the change of environment for agent $i$, \ie $\sum_{i\myin\mathcal{N}} \Delta_i(T) =\Delta(T)$.
\end{itemize}

We have:{\small
\begin{align}
    &\sum_{t=1}^{T} f_t(\,\solopt_t\,) \leq \sum_{t=1}^{T} f_t(\,\solopt_t\cup\solosg_t\,)\label{aux1:1} \\\label{aux1:2}
    &= \sum_{t=1}^{T} f_t(\,\solosg_t\,) + \sum_{t=1}^{T} \sum_{i\myin\calN} f_t(a_{i,\,t}^\opt\,|\,\solopt_{i-1,\,t}\cup\solosg_t\,)\\\label{aux1:3}
    &\leq \sum_{t=1}^{T} f_t(\,\solosg_t\,) + \sum_{t=1}^{T} \sum_{i\myin\calN} f_t(a_{i,\,t}^\opt\,|\,\solosg_{i-1,\,t}\,)\\
    &= 2\sum_{t=1}^{T} f_t(\,\solosg_t\,) \nonumber\\\label{aux1:4}
    &\quad +\sum_{t=1}^{T} \sum_{i\myin\calN} f_t(a_{i,\,t}^\opt\,|\,\solosg_{i-1,\,t}\,) - f_t(a_{i,\,t}^\onalg\,|\,\solosg_{i-1,\,t}\,)\\
    \label{aux1:5}
    &= 2\sum_{t=1}^{T} f_t(\,\solosg_t\,) + \sum_{t=1}^{T} \sum_{i\myin\calN} r_{a^\opt,\,t}^{\hspace{.5pt}(i)} - r_{a^\onalg,\,t}^{\hspace{.5pt}(i)},
\end{align}}where \cref{aux1:1} holds from the monotonicity of $f_t$; \cref{aux1:2,aux1:4} are proved by telescoping the sums; \cref{aux1:3} holds from the submodularity of $f_t$; and \cref{aux1:5} holds from the definition of $r_{a,\,t}^{\hspace{.5pt}(i)}$ (\onalg's line 14). We now complete the proof:
{\small
\begin{align}
    \hspace{-2.8cm}\mathbb{E}\Bigl[{\scenario{Tracking}\text{-}\scenario{Regret}_T^{(1/2)}}(\,\solosg\,)\Bigr] \nonumber                             \end{align}
    \begin{align}\label{aux2:1}
    &= \mathbb{E}\Bigl[\frac{1}{2}\sum_{t=1}^{T} f_t(\,\solopt_t\,) - \sum_{t=1}^{T}f_t(\,\solosg_t\,)\Bigr] \\
    \label{aux2:2}
    &\leq \frac{1}{2}\sum_{t=1}^{T}  \sum_{i\myin\calN} \mathbb{E} \Big[r_{a^\opt,\,t}^{\hspace{.5pt}(i)} - r_{a^\onalg,\,t}^{\hspace{.5pt}(i)}\Big] \\
    \label{aux2:3}
    &= \frac{1}{2}\sum_{t=1}^{T}  \sum_{i\myin\calN} r_{a^\opt,\,t}^{\hspace{.5pt}(i)} - r_{t}^{\hspace{.5pt}(i)\top}\distfsf_t^{(i)} \\ \label{aux2:4}
    &\leq \frac{1}{2}\times 8 \sum_{i\myin\calN} \sqrt{T\Big((\Delta_i(T) +1)\log{(|\calV_i|T)}+l\Big)}\\
    \label{aux2:5}
    &\leq 4 \sqrt{|\calN|T\sum_{i\myin\calN}\Big((\Delta_i(T) +1)\log{(|\calV_i|T)}+l\Big)}\\
    \label{aux2:6}
    &\leq 4 \sqrt{|\calN|T\Big((\Delta(T)+ |\calN|)\log{(\max_{i\myin\calN}{|\calV_i|}T)}+|\calN|l\Big)},
\end{align}}where $l\triangleq\log{(1+\log{(T)})}$, \cref{aux2:1} holds from \Cref{def:tracking_regret}; \cref{aux2:2} holds from \cref{aux1:5}; \cref{aux2:3} holds from the internal randomness of \fsf{i}; {\cref{aux2:4} holds from~\cite[Corollary 1]{matsuoka2021tracking}}; 
\cref{aux2:5} holds from the Cauchy–Schwarz inequality; and \cref{aux2:6} holds since $\sum_{i\myin\calN} \Delta_i(T) =\Delta(T)$. \qed

\subsection{Proof of Monotonicity and Submodularity of Function~\eqref{eq:distance}}\label{app:sub}

It suffices to prove that $\max\colon 2^{\mathbb{R}}\mapsto\mathbb{R}$ is non-decreasing and submodular. 
Indeed, if  $\calA\subseteq \calB\in 2^{\mathbb{R}}$, then $\max(\calA)\leq \max(\calB)$, \ie $\max$ is non-decreasing. Next, consider finite and disjoint $\calB_1\in 2^{\mathbb{R}}$ and $\calB_2\in 2^{\mathbb{R}}$, and an arbitrary real number $s$. Then, using \Cref{tab:app}
we verify that $\max(s\,|\,\calB_1)$ $\geq \max{(s\,|\,\calB_1\cup\calB_2)}$ holds true, \ie $\max$ is submodular. \qed

\input{tab-appendix}

%% file: tab-appendix.tex
\begin{table}[h]
\captionsetup{font=footnotesize}
\centering
\renewcommand{\arraystretch}{1.5}
\begin{tabular}{ |c||c|c |}
\hline
  & $\max{(s\,|\,\calB_1)}$ & $\max{(s\,|\,\calB_1\cup\calB_2)}$ \\
 \hline
 $s\leq\min{(b_1,b_2)}$ & 0 & 0 \\
 \hline
 $b_1<s<b_2$ & $s-b_1$ & 0 \\
 \hline
 $b_2<s<b_1$ & 0 & 0 \\
 \hline
 $\max{(b_1,b_2)}\leq s$ & $s-b_1$ & $s-\max{(b_1,b_2)}$ \\
 \hline
 $b_1=b_2$ & $\max{(0,s-b_1)}$ & $\max{(0,s-b_1)}$ \\
 \hline
\end{tabular}
\renewcommand{\arraystretch}{1}
\caption{Comparison of $\max{(s\,|\,\calB_1)}$ and  $\max{(s\,|\,\calB_1\cup\calB_2)}$ for all relationships of $b_1$, $b_2$, and $s$, where $b_1\triangleq\max{(\calB_1)}$ and $b_2\triangleq\max{(\calB_2)}$.
}
 \label{tab:app}
\end{table}